\documentclass[aip,reprint,jcp,amsmath,amssymb,superscriptaddress]{revtex4-1}
\usepackage{graphicx}
\usepackage{color}
\usepackage[normalem]{ulem}

\newcommand{\lu}{\lambda_\text{u}}
\newcommand{\ls}{\lambda_\text{s}}
\newcommand{\omb}{\omega_\text{b}}

\newcommand{\kT}{k_{\text B}T}

\let\Re\relax
\DeclareMathOperator{\Re}{Re}

\newif\ifHighlitedChanges
\def\ifHighlitedChanges{\iftrue}
\ifHighlitedChanges
  
  \def\STRIKE#1{{\color{red}\sout{#1}}}
\else
  
  \def\STRIKE#1{\relax}
\fi

\begin{document}
\bibliographystyle{apsrev}

\title{Transition state trajectory stability determines barrier crossing rates in chemical reactions induced by time-dependent oscillating fields}
\author{Galen T. Craven}
\affiliation{Center for Computational Molecular Science and Technology, \\
School of Chemistry and Biochemistry, \\
Georgia Institute of Technology, \\
Atlanta, GA  30332-0400}

\author{Thomas Bartsch}
\affiliation{Department of Mathematical Sciences, \\
Loughborough University, \\
Loughborough LE11 3TU, \\
United Kingdom}

\author{Rigoberto Hernandez}
\thanks{Author to whom correspondence should be addressed}
\email{hernandez@chemistry.gatech.edu.}
\affiliation{Center for Computational Molecular Science and Technology, \\
School of Chemistry and Biochemistry, \\
Georgia Institute of Technology, \\
Atlanta, GA  30332-0400}

\begin{abstract}
When a chemical reaction is driven by
an external field, the transition state that the system must pass through  
as it changes from reactant to product%
---for example, an energy barrier---%
becomes time-dependent.
We show that for periodic forcing 
the rate of barrier crossing can be determined
through stability analysis of the non-autonomous transition state.
Specifically, strong agreement is observed between the 
difference in the Floquet exponents describing
stability of the transition state trajectory,
which defines a recrossing-free dividing surface 
[G. T. Craven, T. Bartsch, and R. Hernandez, \textit{Phys. Rev. E} \textbf{89}, 040801(R) (2014)],
and the rates calculated by simulation of ensembles of trajectories.
This result opens the possibility to extract rates
directly from the intrinsic stability of the transition state,
even when it is time-dependent,
without requiring a numerically-expensive simulation of the
long-time dynamics of a large ensemble of trajectories.
\end{abstract}
\maketitle
	
Controlling the rate at which reactants transform to products, 
either to accelerate a chemical process or to bias a reaction toward a certain pathway, 
is fundamental to chemical physics. 
Such kinetic control can be achieved through forcing from an external field,
leading to emergent behavior in molecular structure assembly,\cite{Elsner09,Jager11,Prokop12,Ma13} 
organic synthesis \cite{Lids01}, ultracold chemical reactions \cite{Ni10}
 and single molecule spectroscopy.\cite{Zheng13} 
In these processes, reaction rates are typically obtained through
   transition state theory (TST).\cite{mill93,Truhlar96,hern10a,Peters14}
There are two major obstacles to the implementation of TST.
 First, reactive trajectories must be identified and,
 second, the flux of these reactive trajectories 
though a phase space dividing surface (DS) must be calculated. 
If this DS is recrossed by reactive trajectories, 
TST overestimates the rate. 
Only in cases where this DS is recrossing-free is TST formally exact.

In autonomous systems, the optimal DS is determined by a normally hyperbolic invariant manifold (NHIM).\cite{pollak78,pech79a,deLeon2,Uzer02,hern10a,Ezra2009,Ezra2009a,Teramoto11,Allahem12,Li06prl,Waalkens04b,Waalkens13}
The study of NHIMs is the principle focus of modern reaction dynamics 
in so far as knowledge of their geometry inherently contains the determining 
characteristics of the reaction. 
However, even when a recrossing free DS can be found, 
a rate calculation can be intractable, especially for systems with many degrees of freedom, 
as large numbers of trajectories must be integrated to yield statistically relevant results. 

When a reaction is subjected to a time varying external force,
the geometric structures of TST are known to exist in several cases,
though they become time dependent.\cite{Lehmann00a,Lehmann00b,Lehmann03,Maier01,Dyk04,Dyk05,hern14b}
For chemical reactions that are induced solely by by an external field, 
the coupling of the field with the reacting molecule's dipole moment can 
accelerate the reaction rate,\cite{Miller80}
even for systems that dissipate energy through 
a spontaneous emission process.\cite{Argonov08}

An example of a molecular process where 
an external force influences the transition state geometry,
and thus reaction rates, is the photoinduced isomerization between 
\textit{cis} and \textit{trans} stilbene (Ph-C=C-Ph).\cite{Orlandi79,Waldeck91,Martinez03}
Its unimolecular reaction path can be parameterized through
the torsion angle of the C=C double bond.
Changing the energetics along this path through photoinduction
alters the isomerization reaction rate. 

We show here that when a chemical reaction is periodically
forced by an external field (such as a laser), the reaction rates
are determined directly by the stability of the transition state.
We calculate the reaction rate of a model system
by simulating large ensembles of
trajectories and compare this result with the rate predicted by Floquet analysis
of the transition state trajectory.
Corresponding to the ``chemical method'' 
where the reactant concentration is followed 
as a function of time,\cite{Levinepchem}
we obtain reaction rates from the decay of 
a given initial distribution.
These rates are well-defined because
the decay is exponential 
when averaged over a period of the driving
and independent
of the choice of distribution.
A major result of this work is that 
the rates can be obtained from a Floquet analysis
of the transition state trajectory, 
an unstable periodic orbit close to the barrier top.
This agreement suggests that chemical reaction rates
can be extracted 
directly from the transition state without knowledge
of the dynamics of the reactive population.
This general result could have been anticipated 
from the known connection between the stability of
periodic orbits of Hamiltonian systems and rates,\cite{kadanoff84,skodje90,gaspard98}
but is here established even in the case of
driven systems.

\begin{figure}[t!]
\includegraphics[width=8.5cm,clip]{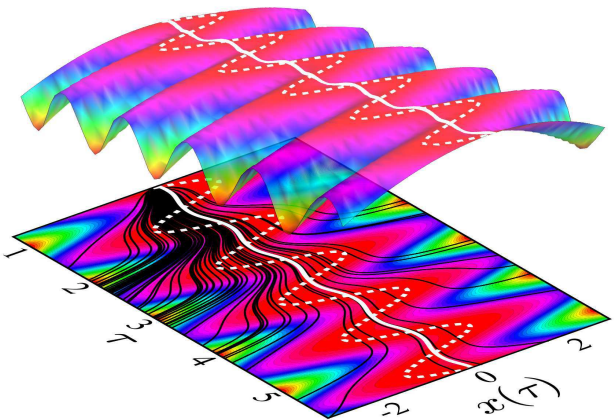}
\caption{\label{fig:Surface}
The time evolution of $x(t)$ for a swarm of trajectories following the Eq.~(\ref{eq:motionAnharm})
with $E(t) = a \sin(\Omega t + \varphi)$ are shown in black (below). 
The potential surface of Eq.~(\ref{eq:potanharm}) is shown above with a contour plot shown below.
The BT and TS trajectories are shown in dashed and solid white, respectively. 
Time is shown in units of 
$\tau = \Omega t / 2 \pi+3/4.$
The initial velocities are sampled from $q_\text{B}$. 
Parameters are $\epsilon = 1$, $\Omega=3$, $\gamma=4$, and $\varphi=0$.}
\end{figure}

To model barrier crossings in chemical reactions driven by a time-dependent external field $E(t)$
we consider a particle of unit mass with an initial position $x_0$ on the reactant side of a moving energy barrier.
The chosen barrier is a quartic potential of the form
\begin{equation}
	\label{eq:potanharm}
	U(x) = -\tfrac 12 \omb^2 (x-E(t))^2-\tfrac{1}{4}\epsilon(x-E(t))^4,
\end{equation}
which leads to the equations of motion
\begin{equation}
\begin{aligned}
	\label{eq:motionAnharm}
	\dot x &= v,  \\
	\dot v &= -\gamma v + \omb^2 (x -  E(t))+\epsilon (x -  E(t))^3,
\end{aligned}
\end{equation}
where $\gamma$ is a dissipative emission parameter, $\omb$ is the barrier frequency, and $\epsilon$ is an anharmonic coefficient.
The anharmonic coefficient is restricted to values $\epsilon\geq0$ 
such that there is a single maximum in the potential located at the barrier top (BT).
The time dependent, instantaneous position of the BT is specified by $E(t)$. 
Figure~\ref{fig:Surface} shows the time evolution of $x(t)$  
for an ensemble of trajectories following Eq.~(\ref{eq:motionAnharm}).
Each trajectory either crosses the energy barrier forming product 
or 
remains on the reactant side, never surmounting the barrier.
The normalized flux of {\it reactive} trajectories through the 
phase-space bottleneck ---the TS---
is the reaction rate.\cite{mill93}

Every realization of the forcing $E(t)$ has 
a special trajectory imbedded in the dynamics (\ref{eq:motionAnharm})
that remains close to the BT for all time. 
This bounded trajectory, termed the transition state (TS) trajectory,\cite{dawn05a,dawn05b,hern06d,Revuelta12,Bartsch12}
will never descend into the product or reactant regions.\cite{hern14b} 
As illustrated in Fig.~\ref{fig:Surface}, the TS trajectory does not follow the time evolution of the energetic maximum given by the BT.
It is instead a specific trajectory that responds to motion of the BT 
in such a way that it remains bounded for all time. 
When $E(t)$ is a periodic function with period $T$
such that $E(t) = E(t+T)$ for all $t$,
the resulting TS trajectory is a periodic orbit (PO) with the same period $T$.

\begin{figure}[t]
\includegraphics[width=8.5cm,clip]{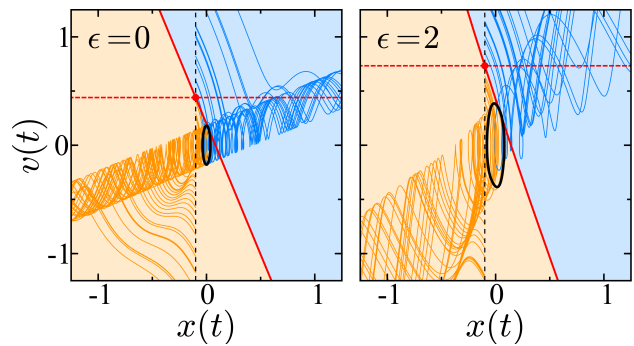}
\caption{\label{fig:Phase}
Phase space plots for a swarm of trajectories following Eq.~(\ref{eq:motionAnharm}) 
with $E(t) = a \sin(\Omega t + \varphi)$.
The initial position for every trajectory, $x_0$, is shown as a dashed black line.
Reactive trajectories are colored in blue and nonreactive trajectories are colored in orange with respective basins
separated by $V^\ddag_c$ (solid red).
The TS~trajectory $\mathbf{\Gamma}^\ddag$ is shown in black.
The critical velocity $V^\ddag$ is indicated by a red circle 
at the intersection of the dashed red line and $x_0$. 
The initial velocities are sampled from $q_\text{B}$. 
Parameters are $\Omega=5$, $\gamma=2$, and $\varphi=0$.}
\end{figure}

Attached to the TS trajectory are stable and unstable manifolds. 
The stable manifold intersects a line of 
initial conditions $x=x_0$ at a critical velocity $V^\ddag$.\cite{Revuelta12,Bartsch12}
A trajectory will surmount the energy barrier, 
moving from the reactant state to the product state, if $v_0>V^\ddag$. 
If $v_0<V^\ddag$, the trajectory is nonreactive.
The extension of this point to all values of $x_0$ creates a critical curve $V^\ddag_c$, 
which is a time-invariant phase space separatrix as illustrated in Fig.~\ref{fig:Phase}.
Knowledge of $V^\ddag_c$ allows the identification of reactive trajectories from initial conditions,
 but it does not contain direct dynamical information such as the reaction rates themselves. 

To calculate rates, the TST methodology is concerned with creating a DS 
that is crossed once and only once by reactive trajectories and then 
evaluating the flux through that surface. 
For the case when $V^\ddag_c$ is known exactly, the no-recrossings criterion is satisfied and 
TST gives the formally exact reaction rate.
In practice, large numbers of trajectories are
generated and the flux is calculated through brute force. 
To construct a recrossing-free DS we will use a time-dependent DS that is located at the instantaneous position of the TS~trajectory.
As shown previously by us,\cite{hern14b} the configuration space projection of the TS trajectory is free of recrossings. 

For the case of a harmonic barrier ($\epsilon=0$), 
Eq.~(\ref{eq:motionAnharm}) can be solved analytically with eigenvalues $\lambda_{\text u,s} = -\frac{1}{2} \left(\gamma \pm \sqrt{\gamma^2 + 4 \omb^2}\right)$ corresponding to the unstable and stable manifolds, respectively.
The TS trajectory is given in Refs.~\citenum{Revuelta12}
and \citenum{Bartsch12} as
\begin{equation}
\begin{aligned}
    \label{eq:TStraj_x}
    x^\ddag(t) &= \frac{\omb^2}{\lu-\ls}\,\left(S[\ls, E;t]-S[\lu, E;t]\right), \\
    v^\ddag(t) &= \frac{\omb^2}{\lu-\ls}\,\left(\ls S[\ls, E;t]-\lu S[\lu, E;t]\right).
\end{aligned}
\end{equation}
in terms of the $S$ functionals\cite{dawn05b,Kawai07}
\begin{equation}
    \label{eq:SDef}
    S_\tau[\mu, g;t] = \begin{cases}
            \displaystyle -\int_t^\infty g(\tau)\,\exp(\mu(t-\tau)) \,d\tau \!\!\!
                & :\; \Re\mu>0, \\[3ex]
            \displaystyle +\int_{-\infty}^t g(\tau)\,\exp(\mu(t-\tau)) \,d\tau \!\!\!
                & :\; \Re\mu<0,
        \end{cases}
\end{equation}
that guarantee the appropriate boundary conditions for $t\to\pm\infty$.
The TS solution for any barrier motion is given by Eq.~(\ref{eq:TStraj_x}).
 
For anharmonic barriers ($\epsilon\ne0$), 
the TS~trajectory will be an unstable PO 
close to the barrier top, as in the harmonic case.
Its period will typically coincide with the period~$T$ of the external driving.
The anharmonic equations of motion (\ref{eq:motionAnharm})
are not amenable to an exact analytical solution,
although approximate analytical methods have previously been employed.\cite{Revuelta12,Bartsch12}
Instead we obtain the TS trajectory
$\mathbf{\Gamma}^\ddag = (x^\ddag(t),v^\ddag(t))$ in phase space numerically
as the periodic solution 
to the system of equations (\ref{eq:motionAnharm}).
A DS that is attached to $\mathbf{\Gamma}^\ddag$ will be recrossing free.
Phase space portraits of $\mathbf{\Gamma}^\ddag$ are shown in Fig.~\ref{fig:Phase}.

\begin{figure}[t]
\includegraphics[width = 8.5cm,clip]{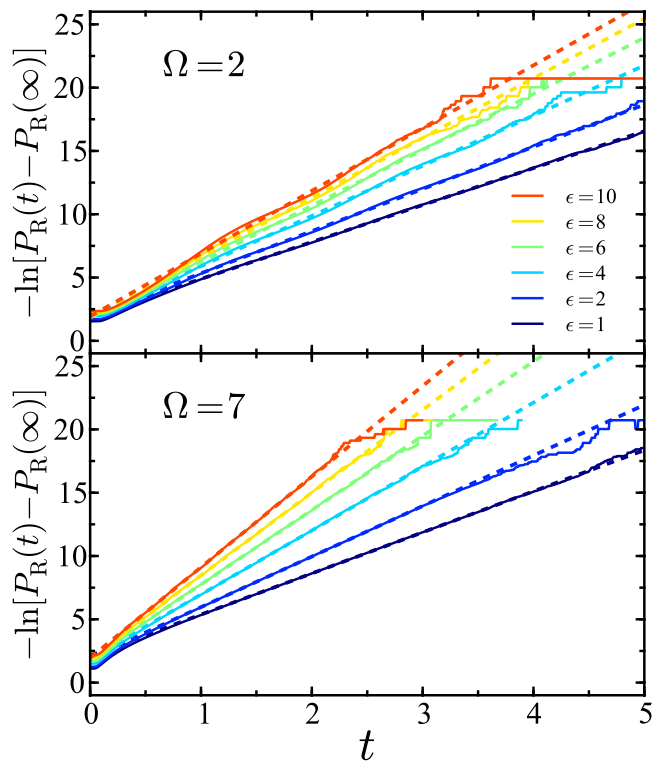}
\caption{\label{fig:Rates}
Time dependence of the scaled logarithm of the reactant population
for $\Omega=2$ and $\Omega=7$ with $v_0$ sampled from $q_\text{B}$.
The slope of each dashed line is the barrier crossing rate $k_\mathrm{f}$, corresponding to a respective $\epsilon$ value.
Parameters are $\gamma=1$, and $\varphi=0$ 
}
\end{figure}

The barrier crossing rates for Eq.~(\ref{eq:potanharm}) were calculated by 
simulating ensembles of trajectories driven by an external field of the form $E(t) = a \sin(\Omega t + \varphi)$.
For single mode sinusoidal driving, the TS trajectory is a PO with period $2 \pi/\Omega$.
Physical units were set by normalizing $a$ and $\omb$ to unity, making all other parameters dimensionless.
Each trajectory was given
an initial position $x_0=-0.1$ to the left of the instantaneous barrier top and
$v_0$ was sampled from two separate distributions: (1) a Boltzmann distribution $q_\text{B}$ with $\kT=1$, and (2) a uniform distribution $q_\text{U}$ (bounded over
the region $\left[V^\ddag - 1/2,V^\ddag + 1/2\right]$). 
For each parameter set 
$\left\{ \Omega,\gamma,\epsilon \right\}$, 
$10^8$-$10^9$ trajectories were simulated.
The normalized reactant population $P_\text{R}(t)$ is obtained from a histogram of
those trajectories that are on the reactant side of the TS~trajectory at time $t$. 
Assuming first order kinetics, the scaled logarithm of 
the normalized population, $-\ln{\left[P_\text{R}(t)-P_\text{R}(\infty)\right]}$,
should be linear in time. 
As illustrated in Fig.~\ref{fig:Rates}, after transient trajectories have crossed, the 
decay of the logarithmic population is linear up to periodic modulation, and the first order assumption is confirmed. 
Periodic fluctuations are noticeable for small driving frequencies ($\Omega \lessapprox 2 $) and large anharmonicities due to effect of
higher order terms in the asymptotic decay of $P_\text{R}(t)$.
The slope of a least squares fit to the non-transient section of the data gives the reaction rates 
calculated from simulation $k_\mathrm{f}$.

We now focus on analysis of the TS trajectory and the determination of reaction rates from its intrinsic stability.
With the bounded TS trajectory now defined, a relative coordinate system can be introduced. 
In relative coordinates
\begin{equation}
	\Delta x = x - x^\ddag(t), \qquad \Delta v = v - v^\ddag(t),
\end{equation}
the equations of motion read
\begin{equation}
\label{eq:motionRel}
\begin{aligned}
	\Delta \dot x &= \Delta v, \\
	\Delta \dot v &= -\gamma \Delta v -U'(\Delta x + x^\ddag(t)) + U'(x^\ddag(t)).
\end{aligned}
\end{equation}
The last term represents a time-dependent driving for the 
relative dynamics that does not depend on the current trajectory. 
It ensures that the relative equations of motion have a fixed point $\Delta \mathbf{\Gamma}^\star$ 
at $\Delta x = \Delta v = 0$, i.e., on the TS~trajectory. 
 
The long-time decay rate of $P_\text{R}(t)$ is determined by the behavior of trajectories close to the stable manifold.
Once a trajectory is sufficiently close to the TS~trajectory, 
it can be described by a linearization of the equations of motion~\eqref{eq:motionRel},
\begin{equation}
\begin{aligned}
\label{eq:motionLin}
	\Delta \dot x &= \Delta v, \\
	\Delta \dot v &= - \gamma \Delta v - a(t)\,\Delta x,
\end{aligned}
\end{equation}
where $a(t) = U''(x^\ddag(t))$. In the phase space vector coordinate $\Delta \mathbf{\Gamma}=(\Delta x,\Delta v)$ this linearization is given by 
\begin{equation}
	\label{eq:gamma_sys}
	\Delta\dot{\mathbf{\Gamma}} =\boldsymbol{J}(t)\,\Delta\mathbf{\Gamma}
\end{equation}
where
\begin{equation}
\label{eq:lin_Jac}
\arraycolsep=5pt\def\arraystretch{1.5}
	\boldsymbol{J}(t)=
	 \begin{pmatrix}
		0 & 1 \\
		\omb^2 +3 \epsilon(x^\ddag(t)-E(t))^2  & -\gamma
	 \end{pmatrix}
\end{equation}
is the Jacobian of Eq.~(\ref{eq:motionRel}) about $\Delta \mathbf{\Gamma}^\star$. 
The linearity of Eq.~\eqref{eq:gamma_sys} 
allows its solution to be expressed as
\begin{equation}
	\Delta\mathbf{\Gamma}(t) = \boldsymbol{\sigma}(t)\, \Delta\mathbf{\Gamma}(\tau)
\end{equation}
where the fundamental matrix solution $\boldsymbol{\sigma}(t)$
is a $2\times 2$ matrix that satisfies
\begin{equation}
	\label{eq:ep_sys}
	\dot{\boldsymbol{\sigma}} =\boldsymbol{J}(t)\,\boldsymbol{\sigma},
	\quad  \boldsymbol{\sigma}(0)=\boldsymbol{I},
\end{equation}
where $\boldsymbol{I}$ is the identity matrix.

The fundamental matrix for one period of~$\Delta \mathbf{\Gamma}^\ddag$
is the monodromy matrix $\boldsymbol{M}=\boldsymbol{\sigma}(T)$
whose eigenvalues $m_{\text u,s}$ are called Floquet multipliers.
The Floquet exponents $\mu_{\text u,s} =1/T \ln|m_{\text u,s}|$ 
give the rates by which nearby trajectories approach or recede from~$\Delta \mathbf{\Gamma}^\ddag$.\cite{chaosbook}
For a harmonic barrier, the multipliers are bonded according to
$0<m_\text{s}<1<m_\text{u}$ giving rise 
to a positive Floquet exponent $\mu_\text{u}$ and a negative exponent $\mu_\text{s}$.
We will assume that this qualitative condition is also satisfied for the anharmonic barriers;
we neglect the possibility that for strong anharmonicities
bifurcations of the TS~trajectory might occur.

Let $\boldsymbol v_\text{u,s}(0)$ be the eigenvectors of $\boldsymbol{M}$.
By Floquet's theorem and the positivity of the Floquet multipliers,
the vectors
\begin{equation}
	\boldsymbol v_\text{u,s}(t) = e^{-\mu_\text{u,s}t}\,\boldsymbol \sigma(t)\,
		\boldsymbol v_\text{u,s}(0)
\end{equation}
are periodic in time with period~$T$.
In the coordinate system defined by these vectors,
\begin{equation}
	\label{eq:oscCoord}
	\Delta\mathbf{\Gamma}(t) = z_\text{u}(t)\,\boldsymbol v_\text{u}(t) + z_\text{s}(t)\,\boldsymbol v_\text{s}(t),
\end{equation}
the linearized equations of motion~\eqref{eq:gamma_sys} read
\begin{equation}
	\dot z_\text{u,s} = \mu_\text{u,s}\,z_\text{u,s},
\end{equation}
with the solution
\begin{equation}
	z_\text{u,s}(t) = C_\text{u,s}\,e^{\mu_\text{u,s}t}.
\end{equation}
Therefore, the vectors $\boldsymbol v_\text{u,s}(t)$ determine the instantaneous directions
of the stable and unstable manifolds in the linear approximation.
The actual stable and unstable manifolds are tangent to these directions at the TS~trajectory.

According to Eq.~\eqref{eq:oscCoord}, the dynamics of Eq.~(\ref{eq:motionLin}) is therefore given by
\begin{equation}
	\label{eq:dxSol}
	\Delta x(t) = C_\text{u}\,\alpha_\text{u}(t)\,e^{\mu_\text{u}t} + C_\text{s}\,\alpha_\text{s}(t)\,e^{\mu_\text{s}t},
\end{equation}
where $\alpha_\text{u,s}$ are the first components of the vectors $\boldsymbol v_\text{u,s}$.
They are periodic with period~$T$.
A trajectory with given initial conditions $C_\text{u}$ and $C_\text{s}$ will cross the moving dividing surface $\Delta x=0$ at time~$t$ determined by
\begin{equation}
	\label{eq:crossTime}
	e^{(\mu_\text{u}-\mu_\text{s})t} = -\frac{C_\text{s}}{C_\text{u}}\,\frac{\alpha_\text{s}(t)}{\alpha_\text{u}(t)}.
\end{equation}
If the initial condition $C_\text{s}$ is fixed and a trajectory with a certain value of $C_\text{u}$ crosses the moving DS at time~$t$, 
 Eq.~(\ref{eq:crossTime}) shows that a trajectory with initial value $C_\text{u}e^{-(\mu_\text{u}-\mu_\text{s})T}$ will cross at time $t+T$. 
Iteration then leads to the existence of trajectories with initial
values $C_\text{u}e^{-(\mu_\text{u}-\mu_\text{s})nT}$ that cross at time $t+nT$.

\begin{figure}[t]
\includegraphics[width = 8.5cm,clip]{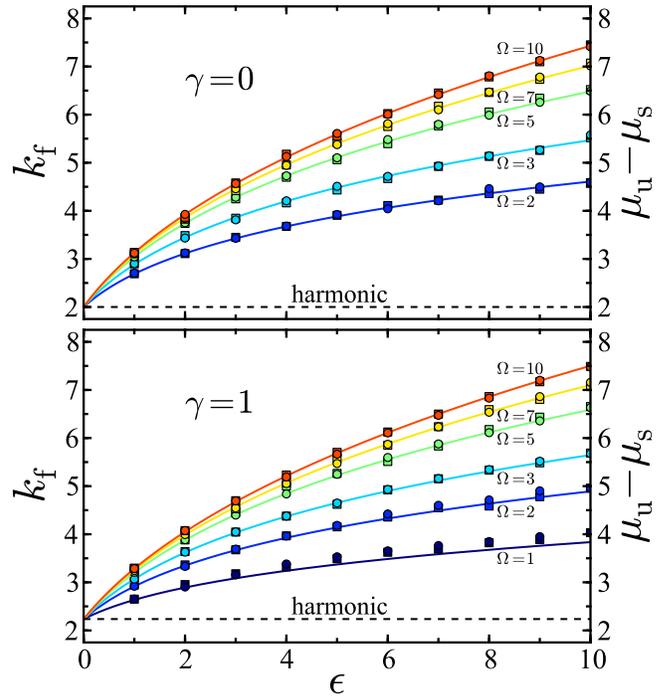}
\caption{\label{fig:Flo}
The barrier crossing rates for Hamiltonian (top) 
and dissipative (bottom) systems following Eq.~(\ref{eq:motionAnharm}).
The numerically calculated rates $k_\mathrm{f}$ for the distributions $q_\text{B}$ (circles) and $q_\text{U}$ (squares)
are shown with units given on the left axes.
The solid curves are the rates predicted by the difference in 
the Floquet exponents $\mu_\text{u}-\mu_\text{s}$ of the TS trajectory
with units given on the right axes.}
\end{figure}

Now consider an arbitrary ensemble of initial conditions with fixed $x(0)$ on the reactant side 
and with a fixed value $C_\text{s}<0$ small enough to be in the region of phase space where the linear approximation (\ref{eq:gamma_sys}) is accurate. 
 In this region the phase space density is constant up to linear corrections in the distance from the stable manifold 
and the number of trajectories that cross the DS in a given time interval is proportional to the width of the strip that contains these trajectories. 
From one period to the next this width decreases by a factor $e^{-(\mu_\text{u}-\mu_\text{s})t}$. 
Thus, up to periodic modulation, the flux must decay by this same factor. 
The flux through the moving DS is the time derivative of the population, $F_\text{M}(t) = \dot{P}(t)$, 
and thus the decay of $P_\text{R}(t)$ is proportional to $e^{-(\mu_\text{u}-\mu_\text{s})t}$. 
From this decay rate it follows that, $k_\mathrm{f} = \mu_\text{u}-\mu_\text{s}$, 
which states that the rate of barrier crossing is the difference in the Floquet exponents.
Note that we have made no assumption for the energy distribution and thus this rate is independent of the ensemble of initial conditions.

A comparison between the rates calculated from numerical simulation $k_\mathrm{f}$, for both the Boltzmann $q_\text{B}$ and uniform $q_\text{U}$ distributions, and rates predicted by the Floquet exponents $\mu_\text{u}-\mu_\text{s}$ is shown in Fig.~\ref{fig:Flo}.
For all values of the forcing frequency $\Omega$, dissipative parameter $\gamma$, and anharmonic strength $\epsilon$, 
the numerical rate is in agreement with rate predicted by stability analysis.
This result opens the possibility that when chemical reactions are forced by periodic external fields 
the reaction rates can be extracted from knowledge of the stability of the TS~trajectory. 
The extension of TS~trajectory stability analysis to aperiodically forced or thermally activated reactions is a focus of our future research.

This work has been partially supported by the National Science Foundation (NSF)
through Grant No.~NSF-CHE-1112067.
Travel between partners was partially supported through the People Programme (Marie Curie Actions)
of the European Union's Seventh Framework Programme FP7/2007-2013/ under REA Grant Agreement No. 294974.

\bibliography{j,hern,gas,tst,osc-bar,halcyon}
\newpage
\printtables
\newpage

\end{document}